\documentclass[%
superscriptaddress,
preprint,
 amsmath,amssymb,
prx,
]{revtex4}
\usepackage{mathptmx}
\usepackage{hyperref}
\usepackage{graphicx}% Include figure files
\usepackage{dcolumn}% Align table columns on decimal point
\usepackage{float}
\usepackage{bm}% bold math
\usepackage[mathlines]{lineno}% Enable numbering of text and display math
\usepackage{color}
\newcommand{\ket}[1]{\left| #1 \right>} % for Dirac bras
\newcommand{\bra}[1]{\left< #1 \right|} % for Dirac kets
\newcommand{\braket}[2]{\left< #1 \vphantom{#2} \right|
 \left. #2 \vphantom{#1} \right>} % for Dirac brackets
\newcommand{\bv}[1]{\ensuremath{\mathbf{#1}}} 
\newcommand{\mathscr}{}

\begin{document}

% \preprint{APS/123-QED}

\title{Electronic structure of bulk manganese oxide and nickel oxide from coupled cluster theory}

\author{Yang Gao}

 \affiliation{Division of Engineering and Applied Science, California Institute of Technology, Pasadena, CA, 91125}
\author{Qiming Sun}
\affiliation{Tencent America LLC, Palo Alto, CA, 94306}
\affiliation{
 Division of Chemistry and Chemical Engineering, California Institute of Technology, Pasadena, CA, 91125
}
\author{Jason M. Yu}
\affiliation{
 Division of Chemistry and Chemical Engineering, California Institute of Technology, Pasadena, CA, 91125
}
\affiliation{Department of Chemistry, University of California, Irvine, CA, 92521}
\author{Mario Motta}%
\affiliation{
 IBM Almaden Research Center, San Jose, CA, 95120
}
\affiliation{
 Division of Chemistry and Chemical Engineering, California Institute of Technology, Pasadena, CA, 91125
}
\author{James McClain}
\author{Alec F. White}
\affiliation{
 Division of Chemistry and Chemical Engineering, California Institute of Technology, Pasadena, CA, 91125
}
\author{Austin J. Minnich}
\email{aminnich@caltech.edu}
 \affiliation{Division of Engineering and Applied Science, California Institute of Technology, Pasadena, CA, 91125}
 
 \author{Garnet Kin-Lic Chan}

\email{gkc1000@gmail.com}
\affiliation{
 Division of Chemistry and Chemical Engineering, California Institute of Technology, Pasadena, CA, 91125
}

\date{\today}

\begin{abstract}
  We describe the ground- and excited-state electronic structure of bulk MnO and NiO, two prototypical correlated electron
  materials, using  coupled cluster theory
  with single and double excitations (CCSD). As a corollary, this work also reports the first implementation
  of unrestricted  periodic ab initio equation-of motion CCSD. 
  Starting from a Hartree-Fock reference, we find fundamental gaps of
  3.46 eV and 4.83 eV for MnO and NiO respectively for the 16 unit supercell, slightly overestimated compared to experiment, although
  finite-size scaling suggests that the gap is more severely overestimated in the thermodynamic limit.
  From the character of the correlated electronic bands we find both MnO and NiO to lie in the 
  intermediate Mott/charge-transfer insulator regime, although NiO appears
  as a charge transfer insulator when only the fundamental gap is considered.
  %% {\color{blue} we see more O p states around close to the Fermi level for NiO, see dos}.
  While the lowest quasiparticle excitations are of metal 3$d$ and O 2$p$ character
  in most of the Brillouin zone, near the $\Gamma$ point, the lowest conduction band quasiparticles are of $s$ character.
  Our study supports the potential of coupled cluster theory to provide high level many-body insights into correlated solids.
%%     We predict both MnO and NiO
%%   to lie in the intermediate region of the 
%%   A spin-unrestricted reference is employed to study the antiferromagnetic (AFM) phase and charged excitations are computed through equation of motion (EOM) formalism.  Our calculations predict MnO to be in the intermediate region of the Zaanen-Sawatzky-Allen
%% phase diagram and NiO to be mostly charge transfer insulator. {\color{red}How do you make such a strong distinction between MnO and NiO from the data?} {\color{blue}Fausely drawn from an old incorrect character analysis(unnormalized), should be changed here. It's hard to distinguish them} The energy gaps are predicted to be 3.46 eV and 4.83 eV for MnO and NiO respectively, a qualitative improvement over mean-field methods. CCSD also exhibits a weak dependence on the reference state. Together, these results indicate that CCSD is promising for an ab-initio description of the electronic structure of moderately correlated solids.
\end{abstract}
\maketitle

\section{\label{sec:intro}Introduction}

%% The accurate first-principles calculation of the electronic structure of

Understanding the properties of solids with correlated electrons is a long-standing challenge.
The first-row transition metal oxides with partially filled $d$ shells, such as MnO and NiO, are prominent examples.
The partially filled $d$ band suggests that these materials should be metals.
However, experimentally they are found to be insulators with large gaps~\cite{sawatzky1984magnitude,van1991electronic,fujimori1990electronic}.
One of the first proposed explanations for this discrepancy was that the gap 
originates from electron interactions, forming a so-called Mott insulator~\cite{mott1949prc}.
Later work, however, based on correlating model cluster calculations to observed spectra~\cite{fujimori1984multielectron,sawatzky1984magnitude} , suggested that the gap
corresponds to a ligand-to-metal charge transfer process.
Since then, the character of the insulating state across the first-row transition-metal oxides has been a fertile topic of study~\cite{wang2012covalency}.

%% Initially, these materials were thought to be prototypical Mott insulators; however this picture was later challenged based on evidence from spectroscopy experiments and model cluster calculation\cite{fujimori1984multielectron,sawatzky1984magnitude}. It was subsequently proposed that the insulating gap is formed by this charge transfer mechanism between the ligand p band and the upper Hubbard band. The later explanation has been widely accepted as it successfully explains most of the photoemission and inverse photoemission data. \cite{sawatzky1984magnitude}
In principle, these questions could be unambiguously resolved through accurate first principles calculation on the bulk material.
However, achieving quantitative accuracy in the properties of greatest interest for transition metal oxides has been difficult.
For example, local and gradient density functional theories (DFT) typically underestimate both the insulating gap and order parameters,
such as the magnetic moment~\cite{terakura1984transition}. While hybrid functionals can give better gaps, this success does not always translate to better properties and is not consistent across materials~\cite{marsman2008hybrid,bredow2000effect, franchini2005density}.
Quantum Monte Carlo methods can provide greater accuracy at higher cost~\cite{ma2015quantum,mitra2015many}, but do not allow access to the full spectrum.
Low-order diagrammatic approaches such as the GW approximation~\cite{faleev2004all,li2005quasiparticle,massidda1997quasiparticle,aryasetiawan1995electronic} have also been applied to these systems, with mixed success. Finally, while DFT with a Hubbard U (DFT+U)~\cite{anisimov1991band,wang2006oxidation} and dynamical mean-field theory (DMFT) calculations~\cite{kunevs2007local,kunevs2007nio,ren2006lda+} have provided a practical approach to obtain important insights, these methods contain a degree of empiricism that introduces uncertainty into the interpretations.

%% The electronic structure of NiO remains a challenge for ab-initio numerical tools. The energy gap and local magnetic moment predicted by standard local spin-density approximation (LSDA) are much smaller than the experimental values\cite{terakura1984transition}. At the density functional theory (DFT) level, "improved" schemes include the generalized gradient approximation (GGA), hybrid functionals, self-interaction corrections (SIC), and $LSDA+U$\cite{leung1991ground,svane1990hydrogen,szotek1993application,anisimov1991band}. how do these perform?[Not sure how to summarize]

%% Beyond the DFT level,  Green's function approaches such as GW approximation and the self-consistent GW method have also been applied to study quasiparticle excitations of NiO with mixed success\cite{faleev2004all,li2005quasiparticle}. For instance, ....examples . Additionally, LDA plus dynamical mean field theory (DMFT) has also been applied to model Hamiltonian of NiO\cite{kunevs2007local,kunevs2007nio,ren2006lda+}. Despite all these studies, a fully ab initio scheme with a systematically improvable framework and no reference dependence is still absent. 

%% Ab initio quantum chemistry community has developed systematically improvable methods based on time-independent perturbation theory\cite{schirmer1983new,neese2003spectroscopy,shavitt2009many}. In this framework,

Coupled cluster (CC) theory is a theoretical framework originating in quantum chemistry and nuclear physics~\cite{shavitt2009many,bartlett2007coupled,vcivzek1966correlation}, which has recently emerged
as a new way to treat electronic structure in solids at the many-body level~\cite{mcclain2017gaussian,gruber2018applying}.
The method is systematically improvable in terms of particle-hole excitation levels, giving rise to the coupled cluster with singles, doubles, triples and higher approximations. While the earliest formulation was for ground states, excited states can be computed via the equation of motion (EOM) formalism~\cite{shavitt2009many,monkhorst1977calculation,krylov2008equation,stanton1993}. Recent single-particle spectra computed for the electron gas~\cite{mcclain2016spectral},
and simple covalent solids~\cite{mcclain2017gaussian} demonstrate that high accuracies can be achieved
at the level of coupled cluster singles and doubles (CCSD). 
%% For systems with weak to moderate correlations, coupled cluster theory with the singles, doubles, and perturbative triples are established as the quantitative “gold standard” of quantum chemistry.

%% Recently, the first excited state CC calculation of a periodic solid was reported, and good accuracy for quantities such as energy gaps was achieved for weakly correlated covalent crystals including silicon and diamond\cite{mcclain2017gaussian}. However, these calculations were performed with a restricted reference, and how well the unrestricted theory performs on prototypical solids like transition metal mono-oxides has not yet been explored.

In this work, we use coupled cluster theory, in its singles and doubles approximation (CCSD and EOM-CCSD), to describe
the ground and excited states of the prototypical transition metal oxides, MnO and NiO. Since
the ground states are magnetically ordered, we use the unrestricted CC formalism, and our work
also reports  the first implementation of unrestricted EOM-CCSD in a periodic system.
%% We compute both ground-state properties
%% as well as the correlated single-particle spectra and estimate the quantitative accuracy that we achieve.
Our largest calculations treat a 2$\times$2$\times$2 supercell of the antiferromagnetic unit cell, with 16 metals and 16 oxygens,
correlating up to 384 electrons. We analyze both the ground state and the correlated single-particle spectra
to report on the character of the insulating state of MnO and NiO.
As we shall see, the  treatment by coupled cluster theory provides an
independent high-level benchmark for many properties and yields new insights into the physics of these materials.

%% to include a spin unrestricted reference to study the ground state and charged excitations of MnO and NiO with AFM order. With GTH-pseudopotential and DZVP basis, our largest calculation amounts to 384 electrons in 1248 orbitals.  CCSD achieves qualitatively improved energy gaps over mean field methods and successfully reproduces the key features of the quasiparticle excitations of the two materials. We also examine the influence of the reference state and perform partial extrapolation to the complete basis set and thermodynamics limit; a complete extrapolation is presently challenging owing to computational limitations.

 The rest of the paper is organized as follows. In Sec.~\ref{sec:theory} we recapitulate the ground state coupled cluster theory and equation of motion (EOM) formalism for excited states in periodic systems. In Sec.~\ref{sec:results} we present the CCSD calculations on NiO and MnO, together
 with analysis of the numerical convergence and character of the states. We finish with some conclusions in Sec.~\ref{sec:conclusion}.

 \section{Methods} \label{sec:theory}

 \subsection{Periodic ground-state coupled cluster theory}

In the coupled cluster formalism, the ground state is described using an exponential excitation Ansatz,
\begin{equation}
    |\Psi_0\rangle = e^{\hat{T}}|0\rangle
\end{equation}
where $|0\rangle$ is a single-determinant reference state obtained from a mean-field theory such
as Hartree-Fock theory or Kohn-Sham density functional theory, and the excitation operator $\hat{T}$ is a sum over
single, double, triple and higher (particle-hole) excitations,
\begin{equation}
    \hat{T} = \sum_{\mu}\hat{t}_{\mu}
\end{equation}
Here $\hat{t}_{\mu}$ creates a linear combination of $\mu$-particle-$\mu$-hole excitations). In CCSD, the cluster operator
is truncated at singles and doubles level so that $\hat{T} = \hat{t}_1 + \hat{t}_2$. In a system with crystal translational symmetry, the particle
and hole states carry a crystal momentum label, thus
\begin{eqnarray}
  \label{eq:amps}
 \hat{t}_1 &=& \sum^{\prime}_{k_i}\sum_{ia} t^{ak_a}_{ik_i}\hat{E}^{ak_a}_{ik_i}\\
 \hat{t}_2 &=& \frac{1}{4}\sum^{\prime}_{k_ik_ak_jk_b}\sum_{iajb}t^{ak_abk_b}_{ik_ijk_j}\hat{a}^\dag_{ak_a}\hat{a}^\dag_{bk_b} \hat{a}_{ik_i}\hat{a}_{jk_j}
%% \hat{t}_1 = \sum^{\prime}_{k_i}\sum_{ia} t^{ak_a}_{ik_i}E^{ak_a}_{ik_i}\\
%% \hat{t}_2 = \frac{1}{2}\sum^{\prime}_{k_ik_ak_jb}\sum_{iajb}t^{ak_abk_b}_{ik_ijk_j}E^{ak_a}_{ik_i}E^{bk_b}_{jk_j}
\end{eqnarray}
where $i,j,k,l\ldots$ denote occupied (hole) spin-orbital labels, $a,b,c,d\ldots$ denote virtual (particle)
labels, $k_i, k_j \ldots$ denote crystal momenta, and $\hat{E}^{ak_a}_{ik_i} = \hat{a}^\dag_{ak_a} \hat{a}_{ik_i}$. 
The primed sum indicates crystal momentum conservation, ie., $k_a + k_b - k_i - k_j = G$ where ${G}$ is a
reciprocal lattice vector. The excitation amplitudes and ground-state energy $E$ are obtained by solving the Schr\"odinger equation for the similarity transformed Hamiltonian $\bar{H}$~\cite{shavitt2009many,bartlett2007coupled},
\begin{eqnarray}
\bar{H} &=& e^{-\hat{T}}\hat{H}e^{\hat{T}}\\
E &=& \langle 0 | \bar{H} | 0 \rangle\\
0 &=& \langle \mu | \bar{H} | 0 \rangle \quad \forall \mu
\end{eqnarray}

%% We can construct the coupled cluster wavefunction to preserve different symmetries. In the restricted theory, $|0\rangle$ is a singlet ($\langle S^2\rangle =\langle S_z\rangle=0$), and the occupied and virtual orbitals come in pairs. In this case, the excitation
%% operator is the same for both spins and we can define $E^{ak_a}_{ik_i} = \sum_{\sigma} E^{\bar{a}k_{\bar{a}} \sigma}_{\bar{i}k_{\bar{i}} \sigma_i}$, and the orbital summations, and restrict the summations in Eq. \eqref{eq:amps} to be summations over only the spatial orbital labels.
In this work, because of the spin ordering in the ground states of MnO and NiO, we will use the unrestricted
form of coupled cluster theory. Here, $|0\rangle$ is a broken-symmetry mean-field state (an eigenfunction of $\hat{S}_z$ but not necessarily $\hat{S}^2$ as, for
example, in an antiferromagnetic state), and $\hat{t}_1$ and $\hat{t}_2$ preserve $S_z$.
%% In this case, we can write $\hat{t}_1$ and $\hat{t_2}$ as sum
%% of terms which separately conserve $S_z$, i.e.
%% \begin{align}
%%   \hat{t}_1 &=& \hat{t}_1^{\alpha\alpha} + \hat{t}_1^{\beta\beta} \\
%%   \hat{t}_2 &=& \hat{t}_2^{\alpha\alpha\alpha\alpha} + \hat{t}_2^{\beta\beta\beta\beta} + \hat{t}_2^{\alpha\alpha\beta\beta}
%%   \end{align}
%% where e.g. $(\alpha\alpha)$, $(\beta\beta)$ denote the spin labels of the summation indices $ia$.
%% The computational cost of periodic CCSD scales as $O(n^4_kn^2_{occ}n^4_{vir})$ where $n_{occ}$, $n_{vir}$ denote
%% the number of occupied and virtual orbitals, respectively. The largest memory cost is associated with holding the $\hat{t}_2$ amplitudes
%% in memory, which are of size $O(n^3_kn^2_{occ}n^2_{vir})$.
The detailed periodic unrestricted CC ground-state equations are given in the Appendix as an extension of
the molecular equations in Refs.~\cite{stanton1991direct,gauss1995coupled}.
%% {\color{red}Put some citations in this section, at least to Bartlett's book.\color{blue} added}
%% For generalization to spin unrestricted reference, spin quantum number must be conserved in the construction of cluster excitation operator. $t_{\mu}$ are thus partitioned into different independent sub-blocks, specifically two spin-pure excitations and one mixed-spin particle-hole excitations for $\mu=2$ while the others are either zero or can be obtained by permutation symmetry.
%% The explicit form of the amplitudes equations are given in the Appendix. 

\subsection{Periodic equation-of-motion coupled cluster theory}

%% {\color{red}Put some citations in this section, at least to Bartlett's book.\color{blue} added}
Excited state wavefunctions and energies in CC can be obtained through the EOM formalism. This amounts to diagonalizing the
similarity transformed Hamiltonian $\bar{H}$ in the excitation space of interest~\cite{bartlett2007coupled,stanton1993,gauss1995coupled,shavitt2009many}. In this work, we compute single-particle
spectra. At the EOM-CCSD level of approximation, we obtain the ionized ($N-1$) (IP) states by diagonalizing in the space of 1-hole (1h) and 2-hole, 1-particle (2h1p) states, while we compute the the electron attached ($N+1$) (EA) states in the space of 1-particle (1p) and 2-particle, 1-hole (2p1h) states. The $n$-th excited state wavefunction with momentum $k$ $|\Psi^{N\pm 1}_{n,k}\rangle $ is constructed as
\begin{eqnarray}
  \label{eq:eomansatz}
|\Psi^{N\pm 1}_{n,k}\rangle &=& \hat{R}^{\pm}_{n,k}|\Psi_0\rangle \\
\hat{R}^{-}_{n,k} &=& \sum\limits_{i}r_{ik}\hat{a}_{ik} +\frac{1}{2} \sum^{\prime}_{k_bk_ik_j}\sum_{bij}r^{bk_b}_{ik_ijk_j}\hat{E}^{bk_b}_{jk_j}\hat{a}_{ik_i}\\
\hat{R}^{+}_{n,k} &=& \sum\limits_{a}r^{ak}\hat{a}^{\dagger}_{ak} +\frac{1}{2} \sum^{\prime}_{k_ak_bk_j}\sum_{abj}r^{ak_abk_b}_{ik_i}\hat{a}^{\dagger}_{ak_a}\hat{E}^{bk_b}_{jk_j}
\end{eqnarray}
(The $n$ index has been omitted from the r.h.s. for clarity; the distinction between subscripts and superscripts is not material but has been
made for correspondence with the literature).
%% where the $R^-$ operator is used for the ionized states and $R^+$ operator used for the electron attached states.
In this work, we use unrestricted $\hat{R}$ operators where $\hat{R}^+$, $\hat{R}^-$ raise or lower $S_z$ by $1/2$, respectively.
%% Similarly to the ground-state theory, the $R$ operators can be chosen to be restricted or unrestricted; in this work we use unrestricted operators.
In the diagonalization step, the dominant computational operation is the multiplication of $\bar{H}$ onto the vector of the R-amplitudes.
The periodic unrestricted equation-of-motion CC formalism has not previously been reported, and detailed equations are given in the Appendix.

\subsection{Computational Details\label{sec:details}}

%\label{sec:details}
MnO and NiO both crystallize in a rock-salt structure. Below the N\'{e}el temperature, the electrons are spin polarized in
stacked ferromagnetic (111) planes with  alternating spin orientations along the [111] direction. In our calculations, we used supercells that are
multiples of a rhombohedral unit cell with four atoms to host the AFM order. All calculations were performed with the experimental lattice constants at 300 K, i.e. $a = 4.43$ $\mbox{\AA}$ and $a = 4.17 $ $\mbox{\AA}$ for MnO and NiO respectively~\cite{cheetham1983magnetic}.
\begin{figure}[h!]%
    \centering
    \includegraphics[width=0.7\columnwidth]{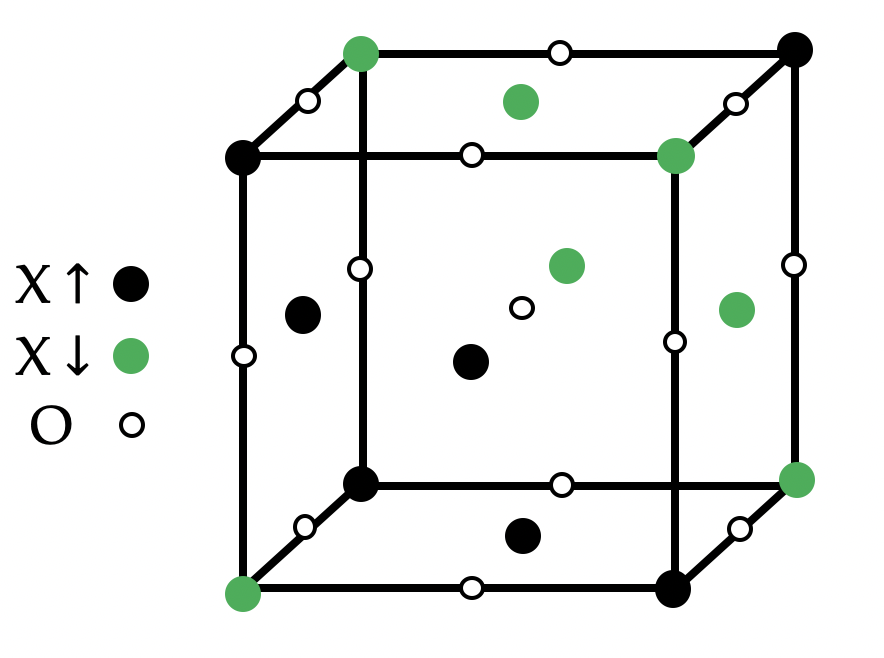}
    \caption{The cell structure of transition metal monoxide XO (X=Mn, Ni) in the AFM II phase. The black and green filled circles denote  metal atoms with
      opposite spin orientations, and the empty circles denote the oxygens.}%
    \label{fig:cell}
\end{figure}

We used the GTH pseudopotential and corresponding single-particle basis~\cite{vandevondele2007gaussian} downloaded from the CP2K package~\cite{VANDEVONDELE2005103}. Most of our calculations used the GTH-DZVP-MOLOPT-SR basis, with 78 orbitals per rhombohedral unit cell; we carried out
a subset of calculations using the GTH-SZV/DZVP/TZVP-MOLOPT(-SR) (SZV/DZVP/TZVP for short) for the metal and oxygen respectively~\cite{vandevondele2007gaussian} to assess
basis set convergence.  Electron repulsion integrals were generated by periodic Gaussian density fitting with an even-tempered
Gaussian auxiliary basis~\cite{sun2017gaussian}.
The initial reference state for the CC calculations was generated from unrestricted Hartree-Fock (UHF) and Kohn-Sham density functional calculations.
Ground state unrestricted CCSD energies and observables used a $2\times 2 \times 2$ Monkhorst-Pack $k$-mesh together with a $2\times 2\times 2$ twist average \cite{gros1992boundary,lin2001twist} to effectively obtain a  $4 \times 4 \times 4$ sampling of the Brillouin zone.
The CC reduced density matrices were computed using only the right eigenvector of $\bar{H}$~\cite{stanton1997ccsd}. Atomic character analysis and local magnetic moments on the metals were computed by population analysis in the crystalline intrinsic atomic orbital basis~\cite{knizia2013intrinsic, cui2019efficient}.
%% DFT band structures were generated in the standard manner, while UHF band structures were generated by
%% twisting a $2\times 2\times 2$ cell and reporting the Hartree-Fock eigenvalues at the (twisted) supercell reciprocal lattice origin.{\color{blue}The DFT/HF band structure is now not shown in any figures, shall we still put the info here.}
Equation-of-motion CCSD band structures were generated using the IP/EA energies at the twisted supercell reciprocal lattice origin.
We sampled these twists along the high-symmetry lines defined by the underlying \textit{non-magnetic primitive cell}
(as generally reported in experiments~\cite{shen1991electronic}). Note that many theoretical calculations report magnetic symmetry labels associated with
the four-atom rhombohedral cell, and we additionally use such labels in parentheses when the high-symmetry points of the primitive and rhombohedral cells coincide.
All methods were implemented within, and calculations performed using, the \textsc{PySCF} package~\cite{sun2018pyscf,PYSCF}. 

%% the left eigenvector of $\bar{H}$  as the right eigenvector {\color{red}there should be a citation to
%%   this approximation.}{\color{blue}I can not find the reference on this one} {\color{red}How are the atomic properties partitioned?} {\color{blue}You mean the local magnetic moment or the character analysis}
%% In the EOM spectral calculations, we centered a uniform k-mesh on the high-symmetry points defined by the underlying face-centered cubic lattice structure. {\color{red}Also specify how the additional points were chosen.}
%% For each high-symmetry point, we used a uniform k-mesh centered at that point for CCSD and subsequent EOM calculations. The full band structure can be computed by performing a denser sampling along the high-symmetry line.

\section{Results\label{sec:results}}

\subsection{Numerical convergence and CC orbital dependence}

\label{sec:convergence}
The CC calculations on these materials are computationally very demanding, and it is not possible to simply calculate properties at the basis
set and thermodynamic limit. We thus first roughly assess the convergence of our numerical results. In Table~\ref{tab:basis} we
report the CC total energy (starting from UHF orbitals), Ni and Mn magnetic moments, and single-particle gap
for a $1\times 1\times 1$ rhombohedral cell (both the direct gap at $\Gamma$ and the presumed fundamental gap that is an indirect transition from $\Lambda_{\frac{1}{2}} (Z)$ (mid-point of the $\Lambda$ symmetry direction of the primitive cell, equivalent to the Z high-symmetry point of the rhombohedral cell, see Fig.~\ref{fig:band} ) to $\Gamma$~\cite{van2006quasiparticle}. Unless otherwise noted, the fundamental gap here will refer to this transition. 
%% Due to the limited sampling of the Brillouin Zone, the fundamental gap might not be exactly from this transition though in all our CC calculations, we found this indirect gap to be the smallest gap. Thus our subsequent discussion on the indirect/fundamental gap will be referring to this specific transition unless otherwise noted.})
As a function of increasing basis size (SZV, DZVP, TZVP),
the magnetic moment is already well converged at the DZVP level, while the total energy still changes
significantly, as expected. The single-particle direct gap $\Delta_\Gamma$ is well converged
while the indirect fundamental gap $\Delta_{ind}$ is somewhat less so, changing by more than 0.1 eV in NiO moving
from DZVP to TZVP. 
%% properties 
%% such as the magnetic moment and single-particle gaps  Indeed the $\Gamma$ point gaps change by less than 0.05 eV{\color{blue}for MnO $\Gamma$ gap changed by 0.09 eV, and $\Lambda_{\frac{1}{2}}\rightarrow\Gamma$ gap is now added to the table. Seems like the there is some basis dependence in the $\Lambda_{\frac{1}{2}}\rightarrow\Gamma$ gap} between DZVP and TZVP.
%% A similar weak basis-dependence of the gap was previously observed in C-diamond, a semiconductor with a comparable bandgap to these materials.
While the remaining basis error may be several tenths of an eV, as a computational compromise, we use the DZVP basis for the remaining calculations. 

\begin{figure}[h!]%
    \centering
     \includegraphics[width=0.8\columnwidth]{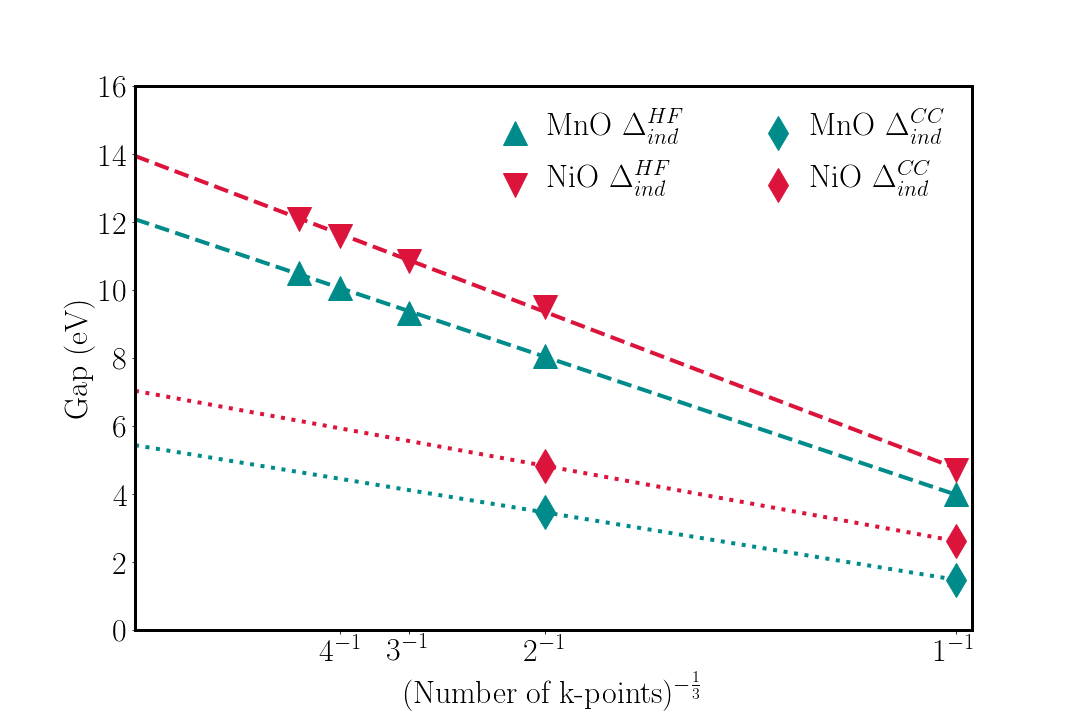}
     \caption{Band gap extrapolation for MnO and NiO. The teal and red triangles denote the HF indirect gap for MnO and NiO respectively. The teal and red diamonds denote the CC indirect gap. The dashed lines and dotted lines give the linear extrapolation to the thermodynamic limit for HF and CC respectively.}%
     \label{fig:hfccgaps}%
 \end{figure}

The same quantities for a $2\times 2\times 2$ supercell (with twist averaging, see Sec.~\ref{sec:details}) are shown in Table~\ref{tab:gs}.
%% {\color{red}Include $\Gamma$ point gap, for direct comparison.}{\color{blue}data added}
Both the magnetic moments and gaps change significantly from the $1\times 1 \times1$ cell; the
change in the indirect and direct gaps ranges from 1.8-2.2 eV. Note that while the basis error converges the gap from above,
the (larger) finite size error converges the gap from below. We can carry out a rough finite-size scaling of the fundamental gap
assuming that it scales as $N_k^{-\frac{1}{3}}$ where $N_k$ is the number of $k$-points in the mesh. This extrapolation is shown in Fig.~\ref{fig:hfccgaps}
for both the UHF and CC gaps. In the TDL, the extrapolated CC gaps increase by a further $~\sim 2$ eV (the precise uncertainty cannot be
gauged from 2 data points).
Taking into account both
finite basis set and size effects, the exact EOM-CCSD gaps in the TDL are thus estimated to be 1-2 eV larger than the $2\times 2\times 2$ results reported here.

%% change the moments by about $10\%${\color{blue} $3\%$ for MnO and $46\%$ for NiO}, and the gaps
%% by about {\color{red}1 eV - correct this once you have the $\Gamma$ point gap in the table}{\color{blue} $\Gamma$ gap changed by 1.77 eV for MnO and 2.34 eV for NiO, $\Lambda_{\frac{1}{2}}\rightarrow\Gamma$ gap changed by 1.98 eV and 2.21 eV for MnO and NiO respectively}.
%% If we naively assume a similar finite size scaling to
%% that seen in C-diamond, then we can expect the gaps in the $2\times 2\times 2$ calculations to be underestimated by {\color{red}0.5-1 eV.} 

%% not needed
\begin{table}[h!]
  \caption{\label{tab:basis}%
    Basis set convergence of CCSD total energy, metal magnetic moment, and direct $\Gamma$ gap $\Delta_\Gamma$ and
  indirect fundamental gap $\Delta_{ind}$  for a $1 \times 1 \times 1$ cell.
}
\begin{ruledtabular}
\begin{tabular}{clcccc}
\textrm{System}&
\textrm{Basis}&
\textrm{$E_{CC}$/eV}&
\textrm{$\mu_B$}&
\textrm{$\Delta^{cc}_{\Gamma}$/eV}&
\textrm{$\Delta^{cc}_{ind}$/eV}\\
\colrule
MnO & SZV    & -2.66 & 4.29 & 0.36 &1.04\\
              & DZVP   & -12.16 & 4.61 & 2.49 &1.48\\
              & TZVP   & -14.23 & 4.61 & 2.40 &1.42\\
\colrule
NiO & SZV    & -3.36 & 0.46 & 2.49 & 2.13\\
              & DZVP   & -13.40 & 1.18 & 3.22 & 2.62\\
              & TZVP   & -15.68 & 1.19 & 3.21 & 2.49\\
\end{tabular}
\end{ruledtabular}
\end{table}

A further variable in the CC calculations is the dependence on the initial mean-field orbitals.
Because of the orbital relaxation generated via the $\hat{t}_1$ operator, this dependence is usually
thought to be mild as long as one is not close to a mean-field symmetry breaking point. In Table~\ref{tab:dft}, we show the $\Gamma$ point
direct gaps for a $1\times 1\times 1$ cell starting from unrestricted DFT orbitals using the LDA~\cite{kohn1965self},
PBE~\cite{perdew1996generalized}, and B3LYP~\cite{becke1993new} functionals,
as well as from unrestricted HF orbitals. (Note that in NiO, only the B3LYP density functional predicts an AFM ground state
when restricted to the $1\times 1\times 1$ cell).
%% {\color{red} doesn't PBE? \color{blue} for $\Gamma$ calculation it can not stablize AFM, 2*2*2 can. I can add that}
%% {\color{red}I think we only need to look at the gaps.}
For MnO, the dependence is very weak, with a maximum variation 
of about 0.16 eV in the gaps. In NiO, the dependence is slightly larger, with a maximum variation of  about 0.47 eV.
This variation is smaller than the combined uncertainties due to basis and finite size effects.

\begin{table}[h!]
  \caption{\label{tab:gs}%
    Local magnetic moment, fundamental gap and direct $\Gamma$ gap from UHF, PBE and CCSD with a $2\times2\times2$ k-point mesh (DZVP basis).
     Extrapolated TDL gap is listed in parentheses. Experimental gaps and moments are also reported (see main text for a discussion of the comparison). The experimental magnetic moment are taken from Refs~\cite{fender1968covalency} and ~\cite{cheetham1983magnetic}. The measured experimental gaps are taken from Refs~\cite{van1991electronic} and ~\cite{sawatzky1984magnitude} for MnO and NiO respectively.}
\begin{ruledtabular}
\begin{tabular}{cccccc}
\textrm{System}&
\textrm{Property}&
\textrm{UHF}&
\textrm{PBE}&
\textrm{CCSD}&
\textrm{exp}\\
\colrule
MnO           & $\mu_{B} $   &  4.86 & 4.56  & 4.76 & 4.58, 4.79\\
             & $\Delta_{ind}$/eV  & 8.05(12.09) & 1.09(1.21)  & 3.46(5.44) &3.6-3.9\\
             & $\Delta_{\Gamma}$/eV & 8.72(13.05) & 1.77(1.84)  & 4.26(5.91) &-\\
\colrule
NiO           & $\mu_{B} $   &  1.85 & 1.34  & 1.72& 1.77, 1.90\\
              & $\Delta_{ind}$/eV  & 9.51(13.95) & 1.19(1.38)  & 4.83(7.04) &4.3\\
              & $\Delta_{\Gamma}$/eV & 9.89(14.80) & 2.45(2.62)  & 5.56(7.90) &-\\
\end{tabular}
\end{ruledtabular}
\end{table}

\begin{table}[h!]
\caption{\label{tab:dft}%
  Direct $\Gamma$ mean-field gaps ($\Delta^{ref}_\Gamma$) and EOM-CCSD gaps ($\Delta^{CC}_\Gamma$) starting from different mean-field orbitals
  for a $1\times 1\times 1$ cell (DZVP basis).
}
\begin{ruledtabular}
\begin{tabular}{clcccc}
\textrm{System}&
\textrm{Method}&
\textrm{IP/eV}&
\textrm{EA/eV}&
\textrm{$\Delta^{ref}_{\Gamma}$/eV}&
\textrm{$\Delta^{CC}_{\Gamma}$/eV}\\
\colrule
MnO, $\Gamma$ & UHF-CC   & -16.77 & -19.26 & 3.99 & 2.49\\
              & LDA-CC  & -16.57 & -19.22 & 1.46 & 2.65\\
              & PBE-CC  & -16.59 & -19.23 & 1.87 & 2.64\\
              & B3LYP-CC  & -16.65 & -19.24 & 2.61 & 2.59\\
\colrule
NiO, $\Gamma$ & UHF-CC   & -18.92 & -22.14 & 4.72 & 3.22\\
              & B3LYP-CC  & -19.27 & -22.02 & 3.63 &2.75
              %% & PBE-CC(PM) & 19.83& 21.67& 1.93&1.84\\
\end{tabular}
\end{ruledtabular}
\end{table}

\subsection{Analysis of the ground state}

We now present a more detailed analysis of the ground states obtained by CCSD for NiO and MnO.
For all results discussed below, we used the unrestricted HF reference.

The CC ground state moments reported in Table~\ref{tab:gs} are significantly reduced from those of UHF,
reflecting the well-known observation that Hartree-Fock tends to overpolarize. Conversely, PBE underpolarizes severely
in NiO.
%% While the CC moment does not appear to have the best agreement with experiment in the case of MnO, 
Note that the theoretical result for the magnetic moment has some variation depending on the definition
of the atomic decomposition, while the experimental error bars are themselves quite large, approximately 0.2 $\mu_B$~\cite{fernandez1998observation}.
Thus the direct comparison between theory and experiment for this quantity should be taken with a degree of caution.

\begin{figure}[h!]%
    %\centering
    \includegraphics[width=0.9\columnwidth]{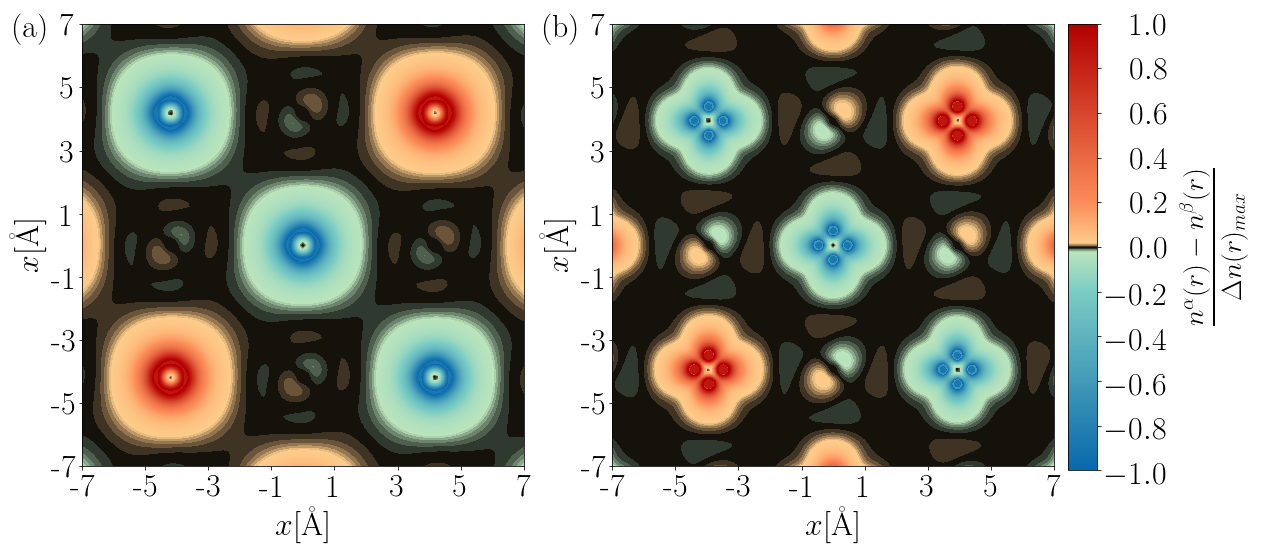}
    \caption{Normalized spin density on the (100) surface for a) MnO and b) NiO. The transition metal atom is located at (0, 0) in the xy-plane. An isotropic spin density on the metal atom is observed for MnO and a clear $e_g$ symmetry is identified for NiO.}%
    \label{fig:rho}%
\end{figure}

Figure~\ref{fig:rho} shows the spin density distribution of the two materials in a (100) surface cut.
For MnO, we find an isotropic spin density, as expected since all 3$d$ orbitals are partially occupied. However, for NiO the
spin density around the Ni atom clearly shows $e_g$ occupancy, while a weakly induced spin density is also observed
around the ligand oxygen site. Note that the O 2$p$ spin density is aligned in the [110] instead of [100] direction, thus allowing maximal superexchange between the nearest Ni sites.

%% Table \ref{tab:gs} summarizes the CCSD ground state properties. As expected,
%% CCSD naturally reduces the magnetic moment relative to that of the HF reference, but is more polarized than DFT.
%% In previous work, the
%% energy difference (per cell) of the FM state and AFM state ($\Delta E_{FM,AFM}$) has been used to parametrize magnetic Hamiltonian exchange couplings {\color{red}include QS-GW work  e.g. by van Schilfgaarde and references}. CCSD provides a significant improvement over the mean-field methods for
%% this quantity, with $\Delta E_{FM,AFM}$ being within 30\% of experiment. {\color{red}This should be modified to report $J$, explaining
%%   how $J$ is obtained from a fictitious FM state energy difference. When the GW results are available, then
%%   a comparison with those is important here. Check: do other people use the smallest cell for this?}{\color{blue} Typically in a magnetic Hamiltonian for this type of system, there are two parameters $J_1$ and $J_2$ for nearest and second nearest exchange coupling. Mapping NiO to a classic model yields $\Delta_{E}=6J_1+6J_2$, we can not extract them both a single equation. Typically people compute another AFM1 phase to get one more equation but we do not have it here. TODO: check}

\begin{figure}[h!]
\includegraphics[width=0.8\columnwidth]{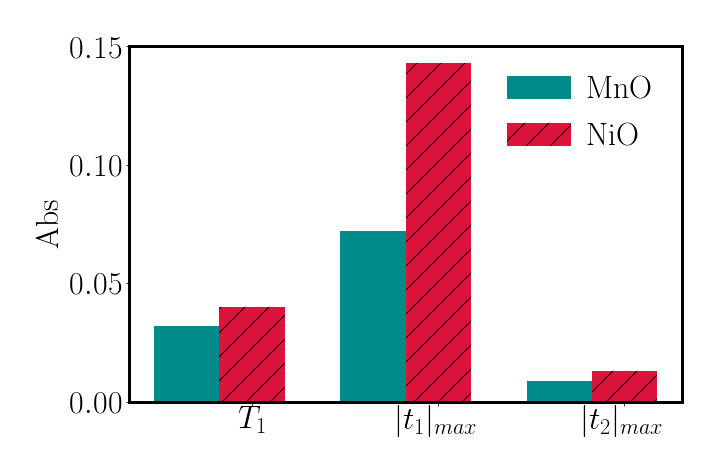}
\caption{\label{fig:amp} CCSD amplitude diagnostics for MnO and NiO. Green columns are for MnO and red are for NiO. $T_1$ is the Frobenius norm of the $t_1$ amplitudes normalized by the number of correlated electrons. %$D_1$ is the maximum matrix 2-norm of $t_1$.
  $|t_1|_{max}$ and $|t_2|_{max}$ are the maximum value for $|t_1|$ and $|t_2|$, respectively.}
\end{figure}

%% While limited single-particle basis set and finite size error could be the potential reason behind the difference between CCSD results and experiment, the inherent multireference nature of the materials could be another source of error.
To further analyze the ground-state correlation, we computed the $T_1$, $|t_1|_{max}$ and $|t_2|_{max}$ diagnostics for the CCSD wavefunction. These are shown in Fig.~\ref{fig:amp}. The $T_1$ metric is the Frobenius norm (normalized by the number of correlated electrons) of the $t_1$ amplitudes.
Previous studies have suggested that values of these diagnostics larger than $\sim 0.1$ can be considered ``large''~\cite{deyonker2007quantitative,lide1995crc}. The $T_1$ and $|t_1|_{max}$ metrics
measure the importance of orbital relaxation from the mean-field reference while $|t_2|_{max}$ measures the true many-particle correlations.
As seen from Fig.~\ref{fig:amp}, the effect of orbital relaxation is greater in NiO than in MnO, consistent with the greater degree of overpolarization
of the Ni moment in the starting HF reference, than is seen for Mn. The small $|t_2|_{max}$ values (0.009 for MnO and 0.013 for NiO) however,
show that both materials are reasonably described by the broken-symmetry mean-field reference.

%% $T_1$ or $|t_1|_{max}$ \textgreater 0.05, $D_1$ \textgreater 0.1 or $D_1$ \textgreater  0.15 \cite{lide1995crc}.
%% The $D_1$ and $|t_1|_{max}$ in our calculations for both materials are well above these criteria, indicating that CCSD is rotating the HF reference. However, the relatively small $T_1$ metric (0.032 for MnO and 0.039 for NiO) and $|t_2|_{max}$ (0.009 for MnO and 0.013 for NiO) shows that the two materials are still far from the strong static correlation region, and therefore that CCSD with a symmetry-broken reference remains a reasonable approximation. Noticeably, through all the metrics that we examined, the values for NiO are larger than MnO by 25\%-98\%, suggesting that NiO is more strongly correlated than MnO.

\subsection{Analysis of the excited states and spectrum}

%% , along with bands obtained from
%% (unrestricted) HF and DFT (PBE){\color{blue} mean field bands are now deleted}.
%% As discussed earlier, we assume that the valence band maximum (VBM) is at $\Lambda_{\frac{1}{2}} (Z)$
%% $[\frac{1}{2} \frac{1}{2} \frac{1}{2}]$ {\color{blue} maybe switch to $\Lambda_{\frac{1}{2}}$?, otherwise it's $\vec{k}=(\frac{1}{2},\frac{1}{2},\frac{1}{2})$ in the unit of reciprocal vector of the cubic unit cell, or $\vec{k}=(\frac{1}{4},\frac{1}{4},\frac{1}{4})$ in the unit of reciprocal vector of the primitive cell with one unit of transtion metal and oxygen}
%% and the conduction band minimum (CBM) is  at the  $\Gamma$.
We next turn to discuss the excited states from EOM-CCSD.
From Table~\ref{tab:gs} we see that the fundamental gaps
obtained by PBE and UHF for MnO are 1.09~eV and 8.05~eV, respectively, both far from the experimental estimate of 3.6-3.9~eV~\cite{van1991electronic}.
In contrast, EOM-CCSD with a $2\times 2\times 2$ supercell finds the indirect gap to be 3.46~eV.
This is similar to the 3.5 eV gap found in prior quasiparticle self-consistent GW (QPscGW) calculations
by Faleev and co-workers~\cite{faleev2004all}. In NiO we observe an indirect gap of 9.51~eV, 1.19~eV and 4.83~eV with HF, PBE and EOM-CCSD ($2\times 2\times 2$ supercell) respectively. The EOM-CCSD gap is much larger than the 2.9 eV gap found by GGA-based GW~\cite{li2005quasiparticle} and close
to the 4.8 eV gap found by QPscGW~\cite{faleev2004all} as well as the experimental estimate of 4.3 eV~\cite{sawatzky1984magnitude}.
However, as discussed in Sec.~\ref{sec:convergence}, the estimated finite size and basis effects in the EOM-CC calculations are quite large
(TDL extrapolations in Table~\ref{tab:gs} are shown
in parentheses) thus the final basis set limit and TDL EOM-CCSD gaps are overestimated by 1-2 eV.
The sizable $T_1$ diagnostics in the ground-state suggest that this error may arise from differential orbital relaxation between the ground
and excited states. Encouragingly, however, we note that the variation in the gap due to
the choice of mean-field orbitals (about 0.5~eV) is significantly smaller than the variation associated with different
starting points in the GW approximation. 

\begin{figure}[htp!]%
    \centering
    \includegraphics[width=0.7\columnwidth]{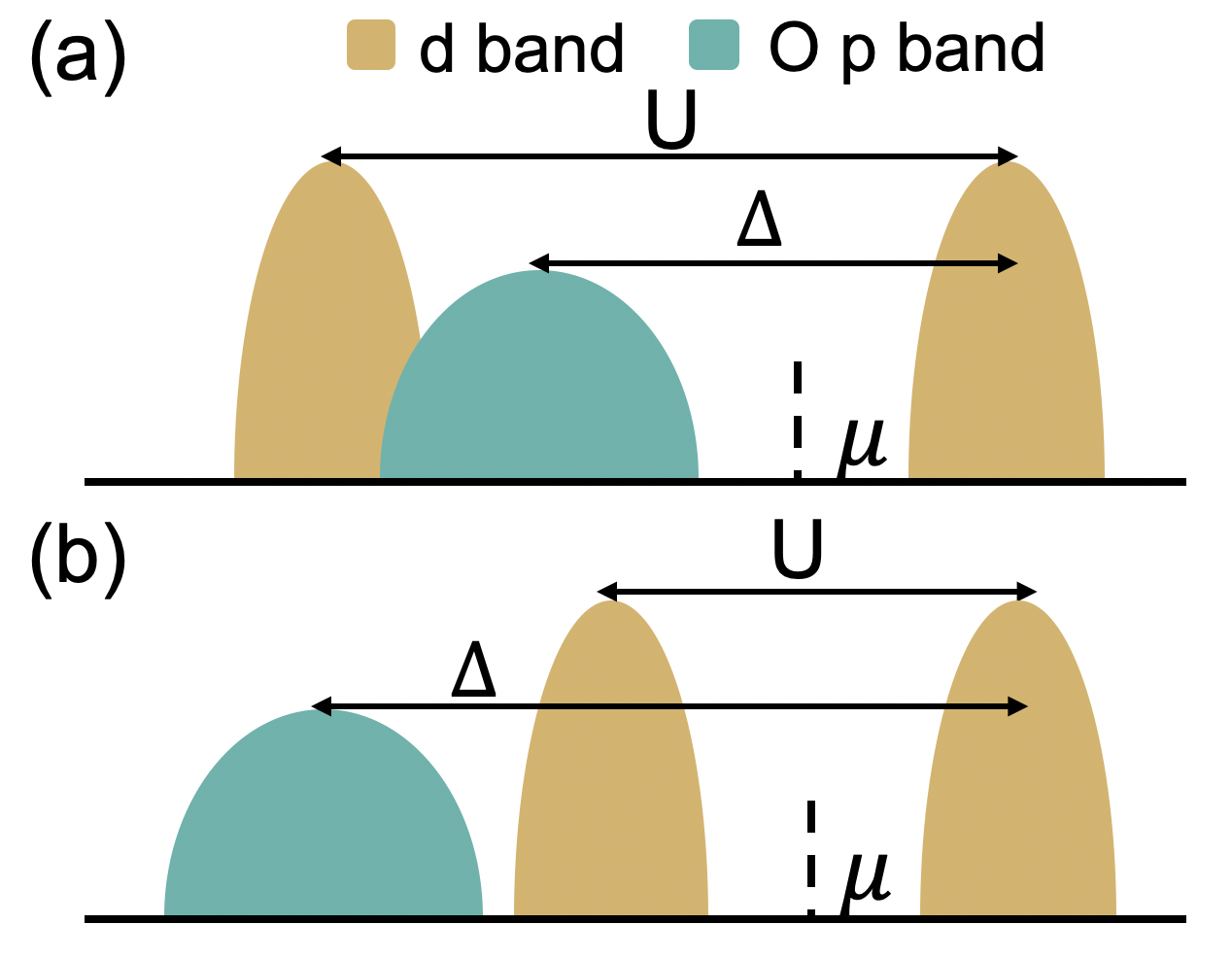}
    \caption{Schematic diagrams of insulating mechanism of a) charge-transfer type where the on-site Coulomb repulsion $U$ is smaller than the charge transfer energy $\Delta$ and b) Mott-Hubbard type where $\Delta>U$. Here $\mu$ is the chemical potential. \label{fig:illustration}}%
\end{figure}

\begin{figure}[htp!]%
    \centering
    \includegraphics[width=0.8\columnwidth]{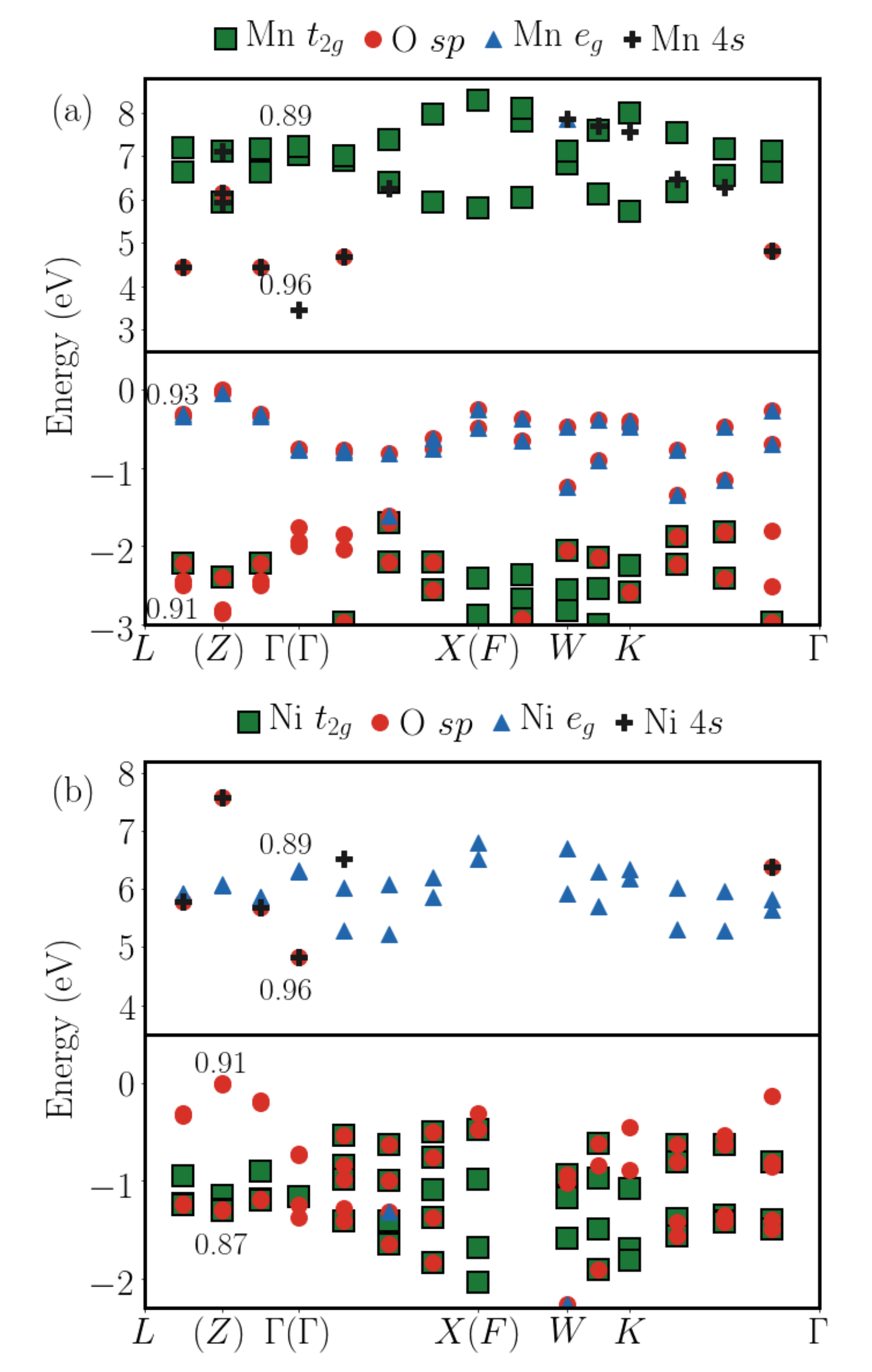}
    \caption{Electronic structure and quasiparticle weight analysis of a) MnO and b) NiO. The labels for the high-symmetry points are those defined by the primitive FCC cell; symmetry labels for the AFM rhombohedral cell are provided in brackets when the special points coincide.
        The upper panel is for the conduction band and the lower one for the valence band. Valence band maxima (VBM) are shifted to 0 eV. Atomic
        character with weight larger than 30\% is indicated by the indicated symbols. Quasiparticle weights are shown for the highest and lowest root computed at $\Gamma$ and $\Lambda_{\frac{1}{2}} (Z)$.  \label{fig:band}}%
\end{figure}

\begin{figure}[htp!]%
    \centering
    \includegraphics[width=0.8\columnwidth]{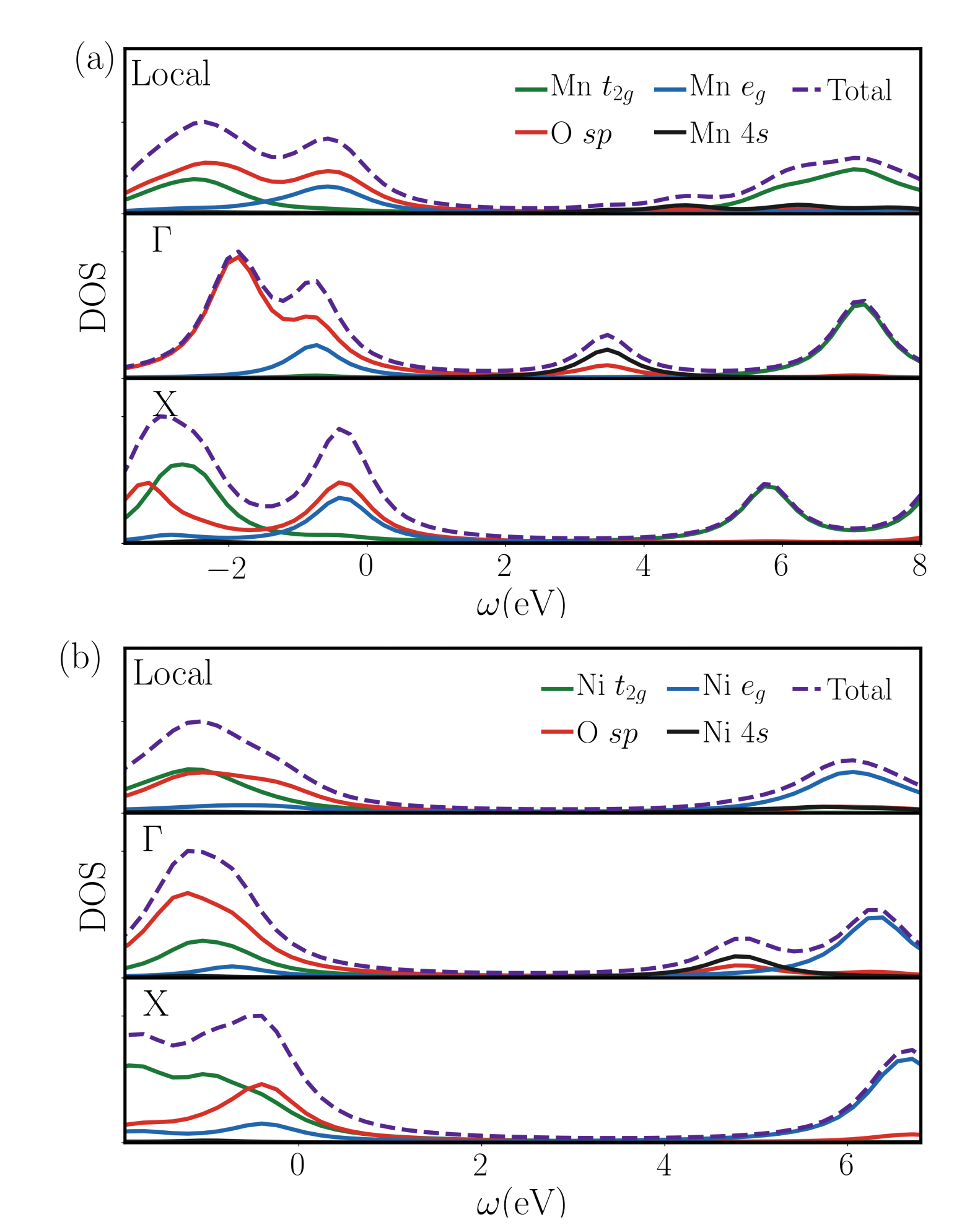}
    \caption{Approximate density of states (DOS) of a) MnO and b) NiO computed by summing over
the EOM-CC roots. The first panel is the local DOS, and the two panels below are the DOS
      at high-symmetry points $\Gamma$ and X. The spectral functions are computed with a Lorentzian broadening factor $\eta=0.4$ eV.%
    \label{fig:dos}}%
\end{figure}

The nature of the insulating gap in MnO and NiO is of some interest. A schematic picture of the charge-transfer and Mott insulating
states is shown in Fig.~\ref{fig:illustration}.
Fig.~\ref{fig:band} plots the correlated band structure at discrete points in reciprocal space from EOM-CC,
with the atomic character labelled by the colours and symbols. Quasiparticle weights are indicated
for selected excitations as the total weight of the 1h (IP) or 1p (EA) sector, i.e. $\sum_{i} |r_{ik}|^2$ and $\sum_{a} |r^{ak}|^2$.
The local density of states at select points in the Brillouin zone (approximated by summing over
the computed EOM-CC roots) is shown in Fig.~\ref{fig:dos}. 

Fig.~\ref{fig:band} shows that the top of the valence band in MnO is hybridized between the $e_g$ states of Mn and the O 2$p$ states,
while the conduction band minimum (CBM) consists mainly of non-dispersive $t_{2g}$ character, except near the $\Gamma$ point (CBM) where
it has $s$ character. In NiO, the valence band near the VBM is dominated by O 2$p$ states (81\% at VBM), while the
picture for the conduction band is similar to that in MnO, including the $s$ character near the CBM.
The above picture is complemented by the DOS in Fig.~\ref{fig:dos} where in MnO, near the Fermi level,
the O 2$p$ states contribute slightly more weight to the valence bands than the Mn $e_{g}$ states, and the two
appear at nearly identical peak positions at around -0.7 eV (relative to the VBM).
The relative positions of the valence $e_{g}$ and $t_{2g}$ bands (-0.7 eV, -2.3 eV) are similar to what is seen in QPscGW (-0.5 eV and -2.2 eV respectively).
%% The peak for the lower $t_{2g}$ band is at around -2.3 eV while the peak for 4s is only visible at $\Gamma$.
Similarly, in NiO, there is little $e_g$ weight (peak around -0.4 eV) near the VBM, and the first peak for $t_{2g}$ is found to be around -1.0 eV. Compared with QPscGW, our calculation suggests less weight for $e_g$ around VBM and the location of $t_{2g}$ is similar to their finding ($\sim$-1.0 eV).
Note that additional valence $e_g$ peaks in NiO are expected to lie deeper in the spectrum~\cite{faleev2004all} and thus do not appear in Fig.~\ref{fig:band}. Quasiparticle weights at the CBM and VBM in both materials are large ($\sim 0.9$).

%% We finish we examine to what extent the excitations near the Fermi surface are renormalized. To do so, we
%% define a quasiparticle weight for each excitation to be the contribution of the 1h or 1p sector in the IP or EA EOM $R$ amplitude, i.e. $\sum_{ik} |r_{ik}|^2$ and $\sum^{ak} |r_{ak}|^2$. Note that this measure of the quasiparticle weight does not include the contribution of $e^T$ in the excited state
%% wavefunction in Eq.~\ref{eq:xxx}, thus it measures the degree of additional correlation in the excited state over
%% that already in the ground-state. {\color{blue} Now we only put the selective data in Fig \ref{fig:band}, so the content/position of this part probably needs to be changed. For both MnO and NiO, the quasiparticle weight dropped slightly away from the lowest excited states. Still, among all the excited states that we computed within Fig \ref{fig:band}, all the quasiparticle weight are larger than 0.80, indicating the validity of Fermi liquid theory for this low energy window}

\begin{figure}[h!]%
    \centering
    \includegraphics[width=0.8\columnwidth]{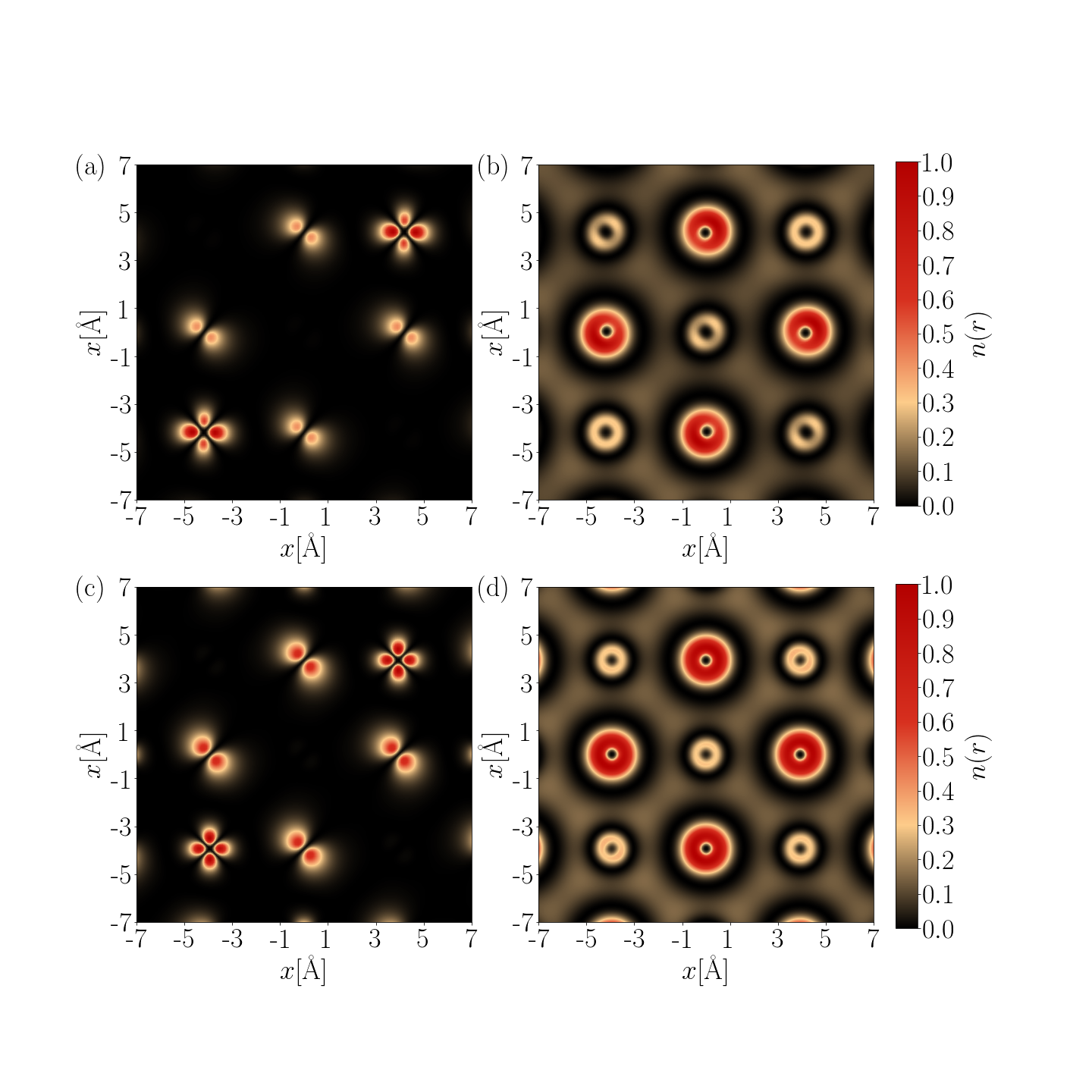}
    %\subfloat[MnO]{{\includegraphics[width=8.5cm]{MnO_ipea_100.png} }}\\%
    %\qquad
    %\subfloat[NiO]{{\includegraphics[width=8.5cm]{NiO_ipea_100.png} }}%
    \caption{Spatial density distribution of quasiparticle orbitals on the (100) surface for (a) MnO VBM, (b) CBM, (c) NiO VBM and (d) CBM. For MnO, we show the $xy$ plane where the projected ionization charge shows $e_g$ symmetry and for NiO, the quasiparticle orbitals are projected onto the $xz$ plane.}%
    \label{fig:wfn}%
\end{figure}

The observed $s$ character of the CBM in MnO and NiO is also found in some earlier GGA-based GW calculations~\cite{li2005quasiparticle},
but not others~\cite{aryasetiawan1995electronic,massidda1997quasiparticle}. This feature is also missed in many DMFT impurity model calculations
where the Ni impurity is defined using only the 3$d$ shell~\cite{kunevs2007local,kunevs2007nio,ren2006lda+} (although a recent full-state ab initio DMFT
calculation on NiO did identify $s$ character at the CBM~\cite{zhu2019efficient}).
The orbital character of the CBM and VBM, including the $s$ character, can be visualized
explicitly in real space by defining quasiparticle orbitals for the CBM/VBM excitation,
\begin{eqnarray}
|\psi_{k}^-\rangle = \sum_{i} r_{ik}|\phi_{ik}\rangle\\
|\psi_{k}^+\rangle = \sum_{a}r^{ak}|\phi_{ak}\rangle
\end{eqnarray}
where $\phi_{ik}$, $\phi_{ak}$ are occupied and virtual mean-field orbitals with crystal momentum $k$.
Real-space density plots of the quasiparticle orbitals at the VBM and CBM are shown in Fig.~\ref{fig:wfn}.

From the analysis above,  both MnO and NiO  appear as insulators of mixed charge-transfer/Mott character.
However, this picture is not uniform across the Brillouin zone. In particular, when only the fundamental gap is examined,
NiO is clearly a charge-transfer insulator while MnO remains of mixed character. Thus the nature of the insulating state in
these systems should be regarded as momentum-dependent.

\section{Conclusion\label{sec:conclusion}}

In conclusion, we have carried out a detailed study of the ground and excited states of MnO and NiO using
coupled cluster theory. While the description of the spectrum is significantly improved over mean-field methods,
and quantitatively accurate at the level of $2\times 2\times 2$ supercells, the gaps in the thermodynamic limit
remain somewhat overestimated, likely due to orbital relaxation effects. Nonetheless,
coupled cluster offers interesting new insights into the qualitative nature of the insulating state in these materials,
allowing for a detailed analysis of the charge-transfer/Mott-insulating character, atomic character of the bands (which indicates
the important participation of $s$ character states in the conduction band minima), and quasiparticle weights. Most intriguingly,
our results show that the charge-transfer Mott nature of the insulating state should be considered to be a momentum-dependent quantity.
Our work is a significant step towards the use of periodic coupled cluster methods to understand correlated electronic
materials. 

\begin{acknowledgments}
  We thank Zhihao Cui and Tianyu Zhu in helpful discussion in spectrum analysis. G. K. C. acknowledges support from DE-SC0018140. Partial support for Y. G. was from DE-SC0019330. A. F. White was supported
  by MURI FA9550-18-1-0095. J.M.Y. acknowledges support from NSF Grant DGE-1745301. Y. G. and A. J. M.  acknowledge the support of ONR under Grant No. N00014-18-1-2101.
\end{acknowledgments}
\clearpage
\appendix

\section{Amplitude Equations}
Here we provide equations for both ground state and IP/EA EOM CCSD starting from an unrestricted mean-field reference.
For compactness, we use bold letters to represent a combined spin-orbital and k-point index i.e. $\bv{i} = (i, k_i)$. Lower-case $i,j,a,b$ and upper-case $I,J,A,B$ are used for spin up and spin down orbitals respectively. The primed sums indicate
momentum conservation. The intermediates $\mathscr{W}$, $\mathscr{F}$, $\tilde{\mathscr{W}}$ and $\tilde{\mathscr{F}}$ that we use
are consistent with those in Stanton and Gauss~\cite{stanton1991direct,gauss1995coupled}. $P(\bv{ab})$ is the antisymmetry operator such that $[P(\bv{ab})Z(\cdot\cdot\cdot ab\cdot\cdot\cdot) = Z(\cdot\cdot\cdot ab\cdot\cdot\cdot)- Z(\cdot\cdot\cdot ba\cdot\cdot\cdot)] $ for any variable Z.

The single excitation amplitudes for up spin obtained from the following equations:
\begin{align}\label{eqn:UT1a}
        0 &= f_{\bv{ai}} + '\sum_{\bv{me}}' t_{\bv{im}}^{\bv{ae}}\tilde{\mathscr{F}}_{\bv{me}}+ \sum_{\bv{ME}}'t_{\bv{iM}}^{\bv{aE}}\tilde{\mathscr{F}}_{\bv{ME}}- \sum_{\bv{mne}}' t_{\bv{mn}}^{\bv{ae}}\braket{\bv{mn}}{\bv{ie}} \nonumber\\
        &-  \sum_{\bv{mNE}}'t_{\bv{mN}}^{\bv{aE}}\braket{\bv{mN}}{\bv{iE}} + \sum_{e}' t_{\bv{i}}^{\bv{e}} \tilde{\mathscr{F}}_{\bv{ae}}- \sum_{m}' t_{\bv{m}}^{\bv{a}}\tilde{\mathscr{F}}_{\bv{mi}} \nonumber\\
        &+ \sum_{\bv{me}}'t_{\bv{me}}\bra{\bv{am}}\ket{\bv{ie}} + \sum_{\bv{ME}}'t_{\bv{M}}^{\bv{E}}\braket{\bv{aM}}{\bv{iE}} \nonumber\\
        &+ \sum_{\bv{mef}}'t_{\bv{im}}^{\bv{ef}}
        \braket{\bv{am}}{\bv{ef}} + \sum_{\bv{MeF}} t_{\bv{iM}}^{\bv{eF}}\braket{\bv{aM}}{\bv{eF}}.
\end{align}
Similarly, the equations for $t^{\bv{A}}_{\bv{I}}$ can be obtained by flipping the spin.

The doubles amplitude equations for $t^{\bv{ab}}_{\bv{ij}}$ are
\begin{align}\label{eqn:U2Taa}
        0 &= \bra{\bv{ab}}\ket{\bv{ij}} + P(\bv{ab}) \sum_{e}' t_{\bv{ij}}^{\bv{ae}}\left[\tilde{\mathscr{F}}_{\bv{be}}- \sum_{\bv{m}}' \frac{1}{2}t_{\bv{m}}^{\bv{b}}\tilde{\mathscr{F}}_{\bv{me}}\right] \nonumber\\
        &- P(\bv{ij}) \sum_{\bv{m}}'t_{\bv{im}}^{\bv{ab}}\left[\tilde{\mathscr{F}}_{\bv{mj}} + \sum_{\bv{e}}' \frac{1}{2}t_{\bv{j}}^{\bv{e}}\tilde{\mathscr{F}}_{\bv{me}}\right] + 
        \sum_{\bv{mn}}' \frac{1}{2}\tilde{\mathscr{W}}_{\bv{mnij}}\tau_{\bv{mn}}^{\bv{ab}}\nonumber \\
        &+   \sum_{\bv{cd}}' \frac{1}{2}\tilde{\mathscr{W}}_{\bv{abcd}}\tau_{\bv{ij}}^{\bv{cd}} + P(\bv{ij})P(\bv{ab})\sum_{\bv{me}}' t_{\bv{i}}^{\bv{e}}t_{\bv{m}}^{\bv{a}}\bra{\bv{mb}}\ket{\bv{je}}\nonumber\\ 
        &+ P(\bv{ij})P(\bv{ab})\left[\sum_{\bv{me}}' t_{\bv{im}}^{\bv{ae}} \tilde{\mathscr{W}}_{\bv{mbej}} + \sum_{\bv{ME}}' t_{\bv{iM}}^{\bv{aE}}\tilde{\mathscr{W}}_{\bv{MbEj}}\right] \nonumber\\
        &+ \sum_{\bv{e}}'P(\bv{ij})t_{\bv{i}}^{\bv{e}}\bra{\bv{ba}}\ket{\bv{je}} - \sum_{\bv{m}}' P(\bv{ab})t_{\bv{m}}^{\bv{a}}\bra{\bv{mb}}\ket{\bv{ij}}.
\end{align}

The equations for $t^{\bv{AB}}_{\bv{IJ}}$ can be obtained by flipping the spins. The equations for $t^{\bv{aB}}_{\bv{iJ}}$ are
\begin{align}
        0 &= \braket{\bv{aB}}{\bv{iJ}} +\sum_{mN}' \tilde{\mathscr{W}}_{\bv{mNiJ}}\tau_{\bv{mN}}^{\bv{aB}} +\sum_{\bv{cD}}' \tilde{\mathscr{W}}_{\bv{aBcD}}\tau_{\bv{iJ}}^{\bv{cD}} \nonumber\\
        &+ \sum_{\bv{E}}' t_{\bv{iJ}}^{\bv{aE}}\left[\tilde{\mathscr{F}}_{\bv{BE}}-\sum_{\bv{M}}'\frac{1}{2}t_{\bv{M}}^{\bv{B}}\tilde{\mathscr{F}}_{\bv{BE}}\right] + \sum_{\bv{e}}' t_{\bv{iJ}}^{\bv{eB}}\left[\tilde{\mathscr{F}}_{\bv{ae}}-\sum_{\bv{m}}'\frac{1}{2}t_{\bv{m}}^{\bv{b}}\tilde{\mathscr{F}}_{\bv{be}}\right] \nonumber\\
        &- \sum_{\bv{M}}'t_{\bv{iM}}^{\bv{aB}}\left[\tilde{\mathscr{F}}_{\bv{MJ}}+\sum_{\bv{E}}'\frac{1}{2}t_{\bv{J}}^{\bv{E}}\tilde{\mathscr{F}}_{\bv{ME}}\right] - \sum_{\bv{m}}'t_{\bv{mJ}}^{\bv{aB}} \left[\tilde{\mathscr{F}}_{\bv{mi}} + \sum_{\bv{e}}'\frac{1}{2}t_{\bv{i}}^{\bv{e}}\tilde{\mathscr{F}}_{\bv{me}}\right] \nonumber \\ 
        &+ \sum_{\bv{me}}'t_{\bv{im}}^{\bv{ae}}\tilde{\mathscr{W}}_{\bv{mBeJ}} +\sum_{\bv{ME}}' t_{\bv{iM}}^{\bv{aE}} \tilde{\mathscr{W}}_{\bv{MBEJ}} - \sum_{\bv{me}}' t_{\bv{i}}^{\bv{e}}t_{\bv{m}}^{\bv{a}}\braket{\bv{eJ}}{\bv{mB}} \nonumber \\ 
        &+ \sum_{\bv{me}}' t_{\bv{mJ}}^{\bv{eB}}\tilde{\mathscr{W}}_{\bv{maei}}
        + \sum_{\bv{ME}}'t_{\bv{MJ}}^{\bv{EB}}\tilde{\mathscr{W}}_{\bv{MaEi}} -\sum_{\bv{ME}}' t_{\bv{J}}^{\bv{E}}t_{\bv{M}}^{\bv{B}}\braket{\bv{aM}}{\bv{iE}} \nonumber \\
        &+ \sum_{\bv{Me}}' t_{\bv{iM}}^{\bv{eB}}\tilde{\mathscr{W}}_{\bv{MaeJ}}  + \sum_{\bv{mE}}'t_{\bv{mJ}}^{\bv{aE}}\tilde{\mathscr{W}}_{\bv{mBEi}}- \sum_{\bv{Me}}'t_{\bv{i}}^{\bv{e}}t_{\bv{M}}^{\bv{B}}\braket{\bv{Ma}}{\bv{Je}} \nonumber \\
        &-\sum_{\bv{mE}}' t_{\bv{J}}^{\bv{E}}t_{\bv{m}}^{\bv{a}}\braket{\bv{mB}}{\bv{iE}} + \sum_{\bv{e}}'t_{\bv{i}}^{\bv{e}}\braket{\bv{Ba}}{\bv{Je}} +\sum_{\bv{E}}' t_{\bv{J}}^{\bv{E}}\braket{\bv{aB}}{\bv{iE}}  \nonumber \\
        &- \sum_{\bv{m}}'t_{\bv{m}}^{\bv{a}}\braket{\bv{iJ}}{\bv{mB}} - \sum_{\bv{M}}' t_{\bv{M}}^{\bv{B}}\braket{\bv{Ji}}{\bv{Ma}}
\end{align}
The remaining mixed spin excitation amplitudes can be obtained via permutational symmetry and need not be computed explicitly.

\begin{align}
    t_{\bv{iJ}}^{\bv{aB}} = t_{\bv{Ji}}^{\bv{Ba}} = -t_{\bv{iJ}}^{\bv{Ba}} = -t_{\bv{Ji}}^{\bv{aB}}
\end{align}

The IP-EOM amplitude equations are given by:
\begin{align}
            (\bar{H}R)_{\bv{i}} &= -\sum_{\bv{k}}' \mathscr{F}_{\bv{ki}} r_{\bv{k}} -  \sum_{\bv{LD}}' \mathscr{F}_{\bv{LD}}r_{\bv{Li}}^{\bv{D}} 
            - \sum_{\bv{ld}}' \mathscr{F}_{\bv{ld}} r_{\bv{li}}^{\bv{d}} \nonumber\\ 
            &- \frac{1}{2} \sum_{\bv{kld}}' \mathscr{W}_{\bv{klid}} r_{\bv{lk}}^{\bv{d}}  - \sum_{\bv{kLD}}' \mathscr{W}_{\bv{kLiD}} r_{\bv{Lk}}^{\bv{D}}
\end{align}
\begin{align}
        (\bar{H}R)_{\bv{Ji}}^{\bv{B}} &= \sum_{\bv{k}}' \mathscr{W}_{\bv{kBiJ}}r_{\bv{k}} - \sum_{\bv{l}}' r_{\bv{Jl}}^{\bv{B}} \mathscr{F}_{\bv{li}} - \sum_{\bv{L}}' r_{\bv{Li}}^{\bv{B}} \mathscr{F}_{\bv{LJ}} \nonumber\\ 
        &+ \sum_{\bv{D}}' r_{\bv{Ji}}^{\bv{D}} \mathscr{F}_{\bv{BD}} + \sum_{\bv{LD}}' \mathscr{W}_{\bv{LBDJ}} r_{\bv{Li}}^{\bv{D}}  + \sum_{\bv{ld}}' \mathscr{W}_{\bv{lBdJ}}r^{\bv{d}}_{\bv{li}}  \nonumber\\
        &+ \sum_{\bv{lD}}' \mathscr{W}_{\bv{lBDi}} r_{\bv{Jl}}^{\bv{D}} + \sum_{\bv{KL}}' \mathscr{W}_{\bv{kLiJ}} r_{\bv{Lk}}^{\bv{B}} \nonumber\\ 
        &- \sum_{\bv{c}}' t_{\bv{iJ}}^{\bv{cB}}\left[\sum_{\bv{kLD}}' \mathscr{W}_{\bv{LkDc}}r_{\bv{Lk}}^{\bv{D}}  + \frac{1}{2} \sum_{\bv{kld}}' \mathscr{W}_{\bv{lkdc}} r^{\bv{d}}_{\bv{lk}}\right]
\end{align}
    
\begin{align}
        (\bar{H}R)^{\bv{b}}_{\bv{ij}} &= - \sum_{\bv{k}}' \mathscr{W}_{\bv{kbji}} r_{\bv{k}} - \sum_{k_l}^{'} \sum_{\bv{l}} \mathscr{F}_{\bv{li}}r^{\bv{b}}_{\bv{jl}} + \sum_{\bv{l}}' \mathscr{F}_{\bv{lj}} r_{\bv{il}}^{\bv{b}}  \nonumber \\
        &+ P(\bv{ij})(\sum_{\bv{LD}}' \mathscr{W}_{\bv{LbDj}}r_{\bv{Li}}^{\bv{D}}  + \sum_{\bv{ld}}'\mathscr{W}_{\bv{lbdj}}r_{\bv{li}}^{\bv{d}}) \nonumber\\  
        & - \sum_{\bv{c}}' t_{\bv{ij}}^{\bv{cb}}\left[\sum_{\bv{kLD}}' \mathscr{W}_{\bv{LkDc}}r_{\bv{Lk}}^{\bv{D}} + \frac{1}{2} \sum_{\bv{kld}}' \mathscr{W}_{\bv{lkdc}} r_{\bv{lk}}^{\bv{d}}\right] \nonumber \\
        &  + \sum_{\bv{d}}' \mathscr{F}_{\bv{bd}}r^{\bv{d}}_{\bv{ji}} +\frac{1}{2} \sum_{\bv{kl}}'\mathscr{W}_{\bv{klij}}r_{\bv{lk}}^{\bv{b}}
\end{align}

The EA-EOM amplitude equations are given by:
 \begin{align}
        (\bar{H}R)^{\bv{a}} &= \sum_{\bv{c}}' \mathscr{F}_{\bv{ac}} r^{\bv{c}} + \sum_{\bv{LD}}' \mathscr{F}_{\bv{LD}}r_{\bv{L}}^{\bv{Da}} + \sum_{\bv{ld}}' 
            \mathscr{F}_{\bv{ld}} r_{\bv{l}}^{\bv{da} } \nonumber \\ 
        &+ \sum_{\bv{cLD}}' \mathscr{W}_{\bv{aLcD}} r_{\bv{L}}^{\bv{Dc}} + \frac{1}{2} \sum_{\bv{cld}}' \mathscr{W}_{\bv{alcd}} r_{\bv{l}}^{\bv{dc}}
\end{align}

\begin{align}
            (\bar{H}R)^{\bv{Ba}}_{\bv{J}} &= \sum_{\bv{c}}' \mathscr{W}_{\bv{aBcJ}} r^{\bv{c}} 
            + \sum_{\bv{c}}' r^{\bv{Bc}}_{\bv{J}} \mathscr{F}_{\bv{ac}} + \sum_{\bv{D}}' r^{\bv{Da}}_{\bv{J}} \mathscr{F}_{\bv{BD}}  \nonumber \\ 
            &- \sum_{\bv{k}}'t_{\bv{kJ}}^{\bv{aB}}  \left[\sum_{\bv{cDL}}' \mathscr{W}_{\bv{kLcD}}r_{\bv{L}}^{\bv{Dc}}  + \frac{1}{2}  \sum_{\bv{cdl}}'\mathscr{W}_{\bv{klcd}}r_{\bv{l}}^{\bv{dc}}\right] 
            \nonumber \\ 
            &+ \sum_{\bv{LD}}' \mathscr{W}_{\bv{LBDJ}}r^{\bv{Da}}_{\bv{L}}  + \sum_{\bv{ld}}' \mathscr{W}_{\bv{lBdJ}}r^{\bv{da}}_{\bv{l}} +  \sum_{\bv{cL}}'\mathscr{W}_{\bv{LacJ}}r^{\bv{Bc}}_{\bv{L}}\nonumber \\
            & - \sum_{\bv{L}}' r_{\bv{L}}^{\bv{Ba}} \mathscr{F}_{\bv{LJ}} + \sum_{\bv{cD}}' \mathscr{W}_{\bv{aBcD}} r_{\bv{J}}^{\bv{Dc}} 
\end{align}
    
\begin{align}
            (\bar{H}R)^{\bv{ba}}_{\bv{j}} &= \sum_{\bv{c}}' \mathscr{W}_{\bv{abcj}}r^{\bv{c}} 
            + \sum_{\bv{c}}' \mathscr{F}_{\bv{ac}}r_{\bv{j}}^{\bv{bc}} + \sum_{\bv{d}}' \mathscr{F}_{\bv{bd}}r^{\bv{da}}_{\bv{j}}  \nonumber \\ 
            & -\sum_{\bv{l}}' \mathscr{F}_{\bv{lj}}r^{\bv{ba}}_{\bv{l}} + \sum_{\bv{cd}}' \frac{1}{2}\mathscr{W}_{\bv{abcd}}r_{\bv{j}}^{\bv{dc}}- \sum_{\bv{LD}}' \mathscr{W}_{\bv{LaDj}}r_{\bv{L}}^{\bv{Db}}\nonumber \\
            &- \sum_{\bv{k}}' t_{\bv{kj}}^{\bv{ab}}\left[\frac{1}{2} \sum_{\bv{cdl}}' \mathscr{W}_{\bv{klcd}}r_{\bv{l}}^{\bv{dc}} + \sum_{\bv{cDL}}' \mathscr{W}_{\bv{LkDc}}r^{\bv{Dc}}_{\bv{L}}\right]  \nonumber \\ 
            &+ \sum_{\bv{ld}}' \mathscr{W}_{\bv{lbdj}}r_{\bv{l}}^{\bv{da}} + \sum_{\bv{LD}}' \mathscr{W}_{\bv{LbDj}}r_{\bv{L}}^{\bv{Da}} - \sum_{\bv{ld}}' \mathscr{W}_{\bv{ladj}}r_{\bv{l}}^{\bv{db}} 
\end{align}

\nocite{*}

\bibliography{ref}% Produces the bibliography via BibTeX.

\end{document}